\documentclass[aps,prd,superscriptaddress,nofootinbib,12pt]{revtex4}

\usepackage{amsmath}
\usepackage{graphicx}

\usepackage{dcolumn}

\usepackage{bm}

\usepackage{amssymb}

\usepackage{latexsym}

\usepackage{subfigure}

\def\ltsima{$\; \buildrel < \over \sim \;$}

\def\gtsima{$\; \buildrel > \over \sim \;$}

\def\simlt{\lower.5ex\hbox{\ltsima}}

\def\simgt{\lower.5ex\hbox{\gtsima}}

\newcommand{\be}{\begin{equation}}

\newcommand{\ee}{\end{equation}}

\newcommand{\bq}{\begin{eqnarray}}

\newcommand{\eq}{\end{eqnarray}}

\newcommand{\pert}{\mathcal{R}}

\begin{document}

\title{Testing a direction-dependent primordial power spectrum with observations of the Cosmic Microwave Background}

\author{Yin-Zhe Ma}

\email{yzm20@cam.ac.uk}

\author{George Efstathiou}

\email{gpe@ast.cam.ac.uk}

\affiliation{Kavli Institute for Cosmology at Cambridge and
Institute of Astronomy, Madingley Road, Cambridge CB3 0HA, United
Kingdom}

\author{Anthony Challinor}

\email{a.d.challinor@ast.cam.ac.uk}

\affiliation{Kavli Institute for Cosmology at Cambridge and
Institute of Astronomy, Madingley Road, Cambridge CB3 0HA, United
Kingdom}

\affiliation{DAMTP, Centre for Mathematical Sciences, Wilberforce
Road, Cambridge CB3 0WA, United Kingdom}

\begin{abstract}
  Statistical isotropy is often assumed in cosmology and should be
  tested rigorously against observational data. We construct simple
  quadratic estimators to reconstruct asymmetry in the primordial
  power spectrum from CMB temperature and polarization data and verify
  their accuracy using simulations with quadrupole power asymmetry. We
  show that the Planck mission, with its millions of signal-dominated
  modes of the temperature anisotropy, should be able to constrain the
  amplitude of any spherical multipole of a scale-invariant quadrupole
  asymmetry at the $0.01$ level ($2\sigma$). Almost
  independent constraints can be obtained from polarization at the
  $0.03$ level after four full-sky surveys, providing an important
  consistency test. If the amplitude of the asymmetry is large enough,
  constraining its scale-dependence should become possible. In
  scale-free quadrupole models with 1\% asymmetry, consistent with the
  current limits from WMAP temperature data (after correction for beam
  asymmetries), Planck should constrain the spectral index $q$ of
  power-law departures from asymmetry to $\Delta q = 0.3$.  Finally,
  we show how to constrain models with axisymmetry in the same
  framework. For scale-free quadrupole models, Planck should constrain
  the direction of the asymmetry to a $1\sigma$  accuracy of about 2 degrees
  using one year of temperature data.
\end{abstract}

\maketitle

\section{Introduction}
\label{sec:intro} In recent years, observations of the cosmic
microwave background (CMB) radiation fluctuations by the Wilkinson
Microwave Anisotropy Probe (WMAP) and a large number of
ground-based and suborbital experiments have led to a precise
measurement of the temperature anisotropy power spectrum up to
multipoles of a few thousand (\cite{Bennett03,Hinshaw07, Brown09,
Reichardt09,ACT10,Hall10}) . Apart from some claimed ``anomalies''
(see below) the observations are consistent with a
dark-energy-dominated cosmology with statistically isotropic,
Gaussian, adiabatic perturbations, as expected from simple models
of inflation. (We will refer to this as the ``concordance''
$\Lambda $CDM model.) If statistical isotropy applies, then the
harmonic coefficients of the temperature field,
\begin{equation}
a_{lm}^{T}=\int d\Omega Y_{lm}^{\ast }(\Omega )\Delta T(\Omega ),
\label{alm-T}
\end{equation}%
must satisfy
\begin{equation}
C_{lm,l^{\prime }m^{\prime }}^{TT}=\left\langle a_{lm}^{T
}a_{l^{\prime }m^{\prime }}^{T\ast}\right\rangle =C_{l}^{TT}\delta
_{ll^{\prime }}\delta _{mm^{\prime }}.  \label{cov-T}
\end{equation}
If the fluctuations are Gaussian and statistically isotropic,
their
statistical properties are completely described by the power spectrum $%
C_{l}^{TT}$.

There have been some hints of ``anomalies'' in the WMAP data,
perhaps suggesting a violation of statistical isotropy. These
include alignments of low-$l$ multipoles
\cite{Tegmark03,Bielewicz04, Land05, Copi06}, evidence for power
asymmetry \cite{Hansen08,Eriksen04} and for a deep cold spot in
the southern Galactic hemisphere \cite{Cruz06,Cruz07}. For the
most part, these anomalies have been found by examining the data
without reference to specific theoretical models. There is,
therefore, an \emph{a posteriori} aspect in computing their
statistical significance which is difficult to assess
\cite{Bennett10}. Some authors have, however, claimed highly
significant discrepancies between the CMB data and the concordance $\Lambda $%
CDM model \cite{Copi10}.

Interest in the CMB anomalies has motivated theorists to build
inflationary models that violate rotational invariance, either via
the addition of a vector field \cite{Ackerman07, Himmetoglu09a,
Himmetoglu09b} or via an isocurvature perturbation
\cite{Erickcek08, Erickcek09}. In addition, a number of
phenomenological models that violate statistical isotropy have
been proposed, which can be tested against observations (e.g. \cite%
{Gordon05, Dvorkin07, Ackerman07}).

Reference~\cite{Ackerman07} considers the phenomenological model
in which the primordial power spectrum depends on a preferred
direction,
\begin{equation}
\mathcal{P}(\mathbf{k})=\mathcal{P}(k)[1+g(k)(\mathbf{k\cdot
n})^{2}],  \label{ACW}
\end{equation}%
where $g(k)$ is some arbitrary function of wavenumber. There has
been considerable interest in this model recently. The authors of \cite%
{Groeneboom08} applied Gibbs sampling to the WMAP five-year maps
to test models with $g(k)=g_* = \mathrm{constant}$, finding strong
evidence for a preferred direction with $g_*=0.15 \pm 0.039$ using
multipoles $l \le 400$. Hanson and Lewis \cite{Hanson09} corrected
some algebraic errors in the analysis of \cite{Groeneboom08} and
applied a simpler quadratic estimator to the WMAP 5-year data.
These authors also found evidence for a highly significant
($9\sigma$) departure from statistical isotropy, but with a
preferred direction suspiciously close to the ecliptic poles,
suggestive of some type of systematic effect in the WMAP data.
This analysis was confirmed by \cite{Groeneboom09}. A subsequent
analysis \cite{Hanson10} showed that asymmetries in the WMAP beams
fully account for the observed violation of statistical isotropy.

Despite this negative conclusion, it is still important to assess
the prospects of constraining violations of statistical isotropy
in the CMB with more precise experiments. In this paper, we focus
on constraints from the Planck satellite, which was launched
successfully in May 2009 and has recently completed its third full
scan of the sky. The Planck satellite has much higher
signal-to-noise ratio in both temperature and polarization than
WMAP \cite{PC05}. It also has higher angular resolution --- 5
arcmin full-width at half-maximum (FWHM) at frequencies $\ge 217
\mathrm{\ GHz}$ --- so asymmetries on scales of the beam width
should have little effect at multipoles $l
\lower.5ex\hbox{\ltsima} 1000$. It is also expected that the
Planck beams will be calibrated to high precision from scans of
bright planets \cite{Huffenberger10}. Following the successful
launch of Planck, the European Space Agency (ESA) has approved a
mission extension until the on-board cryogens are depleted. Planck
is therefore expected to produce almost five sky surveys, compared
to the two sky surveys approved for the nominal mission. This
combination of high sensitivity, high resolution and extended
lifetime allows greater scope for testing for systematic effects
than was possible with WMAP. For example, it becomes possible to
use polarization maps independently of temperature maps to test
for violations of statistical isotropy.

In this paper, we extend the analysis of \cite{Pullen07} to assess
how accurately an extended Planck mission can be used to test
models with an anisotropic primordial power spectrum.
This paper is organized as follows: In Section \ref%
{section-model}, we summarize some basic properties of the
anisotropic model. Section \ref{section-forecast} then applies the
quadratic estimator of \cite{Hanson09}, extended to include
polarization, to compute forecasts for Planck. Our conclusions are
presented in Section \ref{section-conclusion}.

\section{The anisotropic model}
\label{section-model}
\subsection{Covariance matrix}
We write the anisotropic primordial power spectrum as
\begin{equation}
\mathcal{P}(\mathbf{k})=\mathcal{P}(k)\left(1+\sum_{LM}g_{LM}(k)Y_{LM}(\mathbf{\hat{k}})\right).
\label{anisotropy-primordial}
\end{equation}%
We assume that parity-invariance continues to hold in the mean so
that $L$ is restricted to even values such that
$\mathcal{P}(-\mathbf{k})=\mathcal{P}(\mathbf{k})$. In this paper,
we consider a quadrupole modulation, i.e.\ $L=2$, $|M| \leq 2$,
with a power-law scale dependence on the wave number
$g_{LM}(k)=g_{LM}(k_{0}/k)^{q}$ where the pivot scale $k_0=0.002\,
\mathrm{Mpc}^{-1}$.
Scale-invariant modulation corresponds to $q=0$.

The harmonic coefficients of the CMB anisotropy can be expressed
as
\begin{equation}
a_{lm}^{X}=4\pi i^{l}\int \frac{d^{3}\mathbf{k}}{(2\pi
)^{3}}\Delta _{l}^{X}(k) \pert (\mathbf{k})Y_{lm}^{\ast
}(\mathbf{\hat{k}}), \label{alm}
\end{equation}%
where $\Delta _{l}^{X}(k)$ are the adiabatic transfer functions,
either for temperature ($X=T$) or $E$-mode polarization ($X=E$).
The primordial curvature perturbation is $\pert(\mathbf{k})$ with
statistically-homogeneous but anisotropic correlations
\begin{equation}
\left\langle \pert(\mathbf{k})\pert^{\ast }(\mathbf{k}^{\prime
})\right\rangle =(2\pi )^{3}\delta ^{3}(\mathbf{k}-\mathbf{k}^{\prime })%
\frac{2\pi ^{2}}{k^{3}}\mathcal{P}(\mathbf{k}),
\end{equation}%
with $\mathcal{P}(\mathbf{k})$ given by Eq.
(\ref{anisotropy-primordial}).
Thus, the covariance matrix of the harmonic coefficients is%
\begin{eqnarray}
C_{l_{1}m_{1},l_{2}m_{2}}^{XX^{\prime }} &=&\left\langle
a_{l_{1}m_{1}}^{X}a_{l_{2}m_{2}}^{X^{\prime }\ast }\right\rangle  \notag \\
&=&C_{l_{1}}^{XX^{\prime }}\delta _{l_{1}l_{2}}\delta
_{m_{1}m_{2}}+\delta C_{l_{1}m_{1},l_{2}m_{2}}^{XX^{\prime }},
\label{cov-angular}
\end{eqnarray}%
where%
\begin{equation}
C_{l_{1}}^{XX^{\prime }}=4\pi \int d\ln k \mathcal{P}(k)\Delta
_{l_{1}}^{X}(k)\Delta _{l_{1}}^{X^{\prime }}(k)
\end{equation}%
is the usual isotropic power spectrum. The additional term in
Eq.~(\ref{cov-angular}) due to the power asymmetry is
\begin{eqnarray}
\delta C_{l_{1}m_{1},l_{2}m_{2}}^{XX^{\prime }} &=&i^{l_{1}-l_{2}}\tilde{C}%
_{l_{1}l_{2}}^{XX^{\prime }}(q)\sum_{LM}g_{LM}\int d\Omega _{k}Y_{LM}(%
\mathbf{\hat{k}})Y_{l_{1}m_{1}}^{\ast }(\mathbf{\hat{k}})Y_{l_{2}m_{2}}(%
\mathbf{\hat{k}})  \notag \\
&=&i^{l_{1}-l_{2}}\tilde{C}_{l_{1}l_{2}}^{XX^{\prime
}}(q)\sum_{LM} g_{LM}  \notag \\
&&\mbox{} \times(-1)^{m_{1}}\left[ \frac{(2L+1)(2l_{1}+1)(2l_{2}+1)}{4\pi }\right] ^{\frac{1}{2}%
}\left(
\begin{array}{ccc}
L & l_{1} & l_{2} \\
0 & 0 & 0%
\end{array}%
\right) \left(
\begin{array}{ccc}
L & l_{1} & l_{2} \\
M & -m_{1} & m_{2}%
\end{array}%
\right) ,  \label{deltaXX}
\end{eqnarray}%
where
\begin{equation}
\tilde{C}_{l_{1}l_{2}}^{XX^{\prime }}(q)=4\pi \int d\ln k
\mathcal{P}_{\pert }(k)\Delta _{l_{1}}^{X}(k)\Delta
_{l_{2}}^{X^{\prime }}(k)\left( \frac{k_{0}}{k}\right) ^{q}.
\label{cl1}
\end{equation}
Since $L$ is even, the anisotropic covariance is nonzero only for
$l_1-l_2$ even as required by parity invariance. In the following
it will also be convenient to introduce $2\times 2$ matrices
$\mathbf{C}_{l_1 m_1,l_2 m_2}$ and $\tilde{\mathbf{C}}_{l_1 l_2}$
with elements $C_{l_1 m_1,l_2 m_2}^{X_1 X_2}$ and $\tilde{C}_{l_1
l_2}^{X_1 X_2}(q)$ respectively.
\subsection{Quadratic estimators and the Fisher matrix}
\label{section-estimator} Here we assume that the scale-dependence
of the power asymmetry (i.e.\ $q$) is known. We can then use the
quadratic estimator of Ref. \cite{Hanson09}, extended to
polarization, to form estimates $\hat{g}_{LM}$ of the anisotropy
parameters. For an isotropic survey and in the limit of small
primordial anisotropy, these take the form
\begin{equation}
\hat{g}_{LM} = \frac{1}{2} \sum_{L'M'} F^{-1}_{LM,L'M'} \sum_{X_1
l_1 m_1}\sum_{X_2 l_2 m_2} \bar{a}_{l_1 m_1}^{X_1 \ast}
\frac{\partial C_{l_1 m_1 , l_2 m_2}^{X_1 X_2}}{
\partial g_{L' M'}^*} \bar{a}_{l_2 m_2}^{X_2} .
\label{eq:estimator}
\end{equation}
Here, $\bar{a}_{lm}^X \equiv
\sum_{X'}[(\mathbf{C}_l^{\mathrm{tot}}) {}^{-1}]^{XX'}
a^{X'}_{lm}$ are the temperature and polarization multipoles after
weighting with the inverse of their isotropic total
(signal-plus-noise) covariance matrix. The Fisher matrix,
evaluated at $g_{LM}=0$, is given by
\begin{equation}
F_{LM,L'M'} = \frac{1}{2} \sum_{l_1 m_1}\sum_{l_2
m_2}\mathrm{Tr}\left[(\mathbf{C}_{l_1}^{\mathrm{tot}})^{-1}
\frac{\partial \mathbf{C}_{l_1 m_1, l_2 m_2}}{\partial g_{LM}^*}
(\mathbf{C}_{l_2}^{\mathrm{tot}})^{-1} \frac{\partial
\mathbf{C}_{l_2 m_2, l_1 m_1}}{\partial g_{L'M'}} \right] .
\label{eq:Fisher}
\end{equation}
In the limit of vanishing primordial anisotropy, the inverse of
this Fisher matrix equals the covariance of the errors on
$\hat{g}_{LM}$, i.e.\ $F^{-1}_{LM,L'M'} = \langle \hat{g}_{LM}
\hat{g}^\ast_{L'M'}\rangle$.

The assumed isotropy of the survey ensures that the Fisher matrix
at $g_{LM}=0$ is diagonal. Using Eq.~(\ref{deltaXX}), the diagonal
elements evaluate to
\begin{equation}
F_{LM,LM} =\sum_{l_{1}l_{2}}\left[
\text{Tr}\left(\tilde{\mathbf{C}}_{l_{1}l_{2}}
(\mathbf{C}_{l_{2}}^{\mathrm{tot}})^{-1}\tilde{\mathbf{C}}_{l_{2}l_{1}}
(\mathbf{C}_{l_{1}}^{\mathrm{tot}})^{-1}\right)\frac{%
(2l_{1}+1)(2l_{2}+1)}{8\pi }\left(
\begin{array}{ccc}
l_{1} & l_{2} & L \\
0 & 0 & 0%
\end{array}%
\right)^{2}\right] \label{FM2}.
\end{equation}

Since $g_{LM}$ is complex for $M\neq 0$, we present our simulation
results in the next section in terms of $\hat{G}_{LM}=\sqrt{2}{\rm
Re}(\hat{g}_{LM})$ and $\hat{G}_{L-M}=\sqrt{2}{\rm
Im}(\hat{g}_{LM})$ for $M>0$, and $\hat{G}_{L0}=\hat{g}_{L0}$. The
$\hat{G}_{LM}$ are uncorrelated and have the same variance as the
$\hat{g}_{LM}$.
\section{Forecasts for Planck}
\label{section-forecast}
\subsection{Constraints on the anisotropy amplitude}
\label{subsec:amp_cons} In this section, we consider forecasts for
the Planck mission. We use the parameters for the $143 \;
\mathrm{GHz}$ channel (the most sensitive of the Planck frequency
channels) as given in \cite{PC05}. We therefore assume a Gaussian
beam with FWHM of $7.1$ arcmin and assume uncorrelated isotropic
noise in the temperature and polarization maps with
root-mean-square noise levels of $\sigma_T$ and $\sigma_P$
respectively. For the nominal two-sky-survey mission (one year of
observation) we adopt $\sigma _{T}=12.2 \;\mu \mathrm{K}$ and
$\sigma _{P}=23.3 \; \mu \mathrm{K}$ in $3.4$ arcmin pixels
(Healpix~\cite{Gorski05} resolution $N_{\mathrm{side}}=1024$),
corresponding to $42\,\mu\mathrm{K}$-arcmin noise in temperature
and $80\,\mu\mathrm{K}$-arcmin in Stokes $Q$ and $U$ polarization.
We also consider an extended mission of four complete sky surveys
(two years of observation) with $\sigma_T$ and $\sigma_P$ reduced
by a factor of $\sqrt 2$.

\begin{figure}[tbp]
\centerline{\includegraphics[bb=19 141 748
544,width=3.4in]{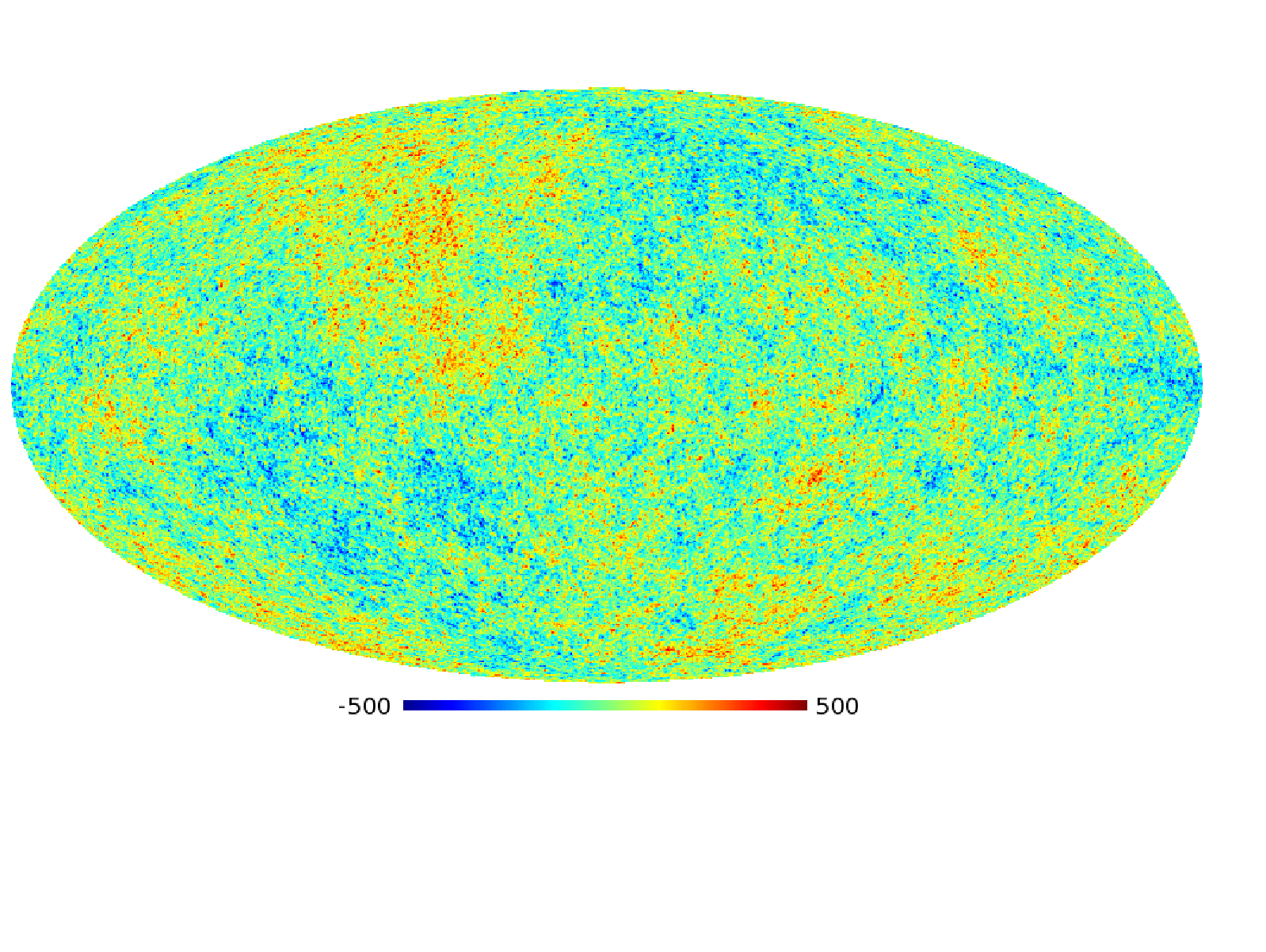}
\includegraphics[bb=19 141 773 548,width=3.4in]{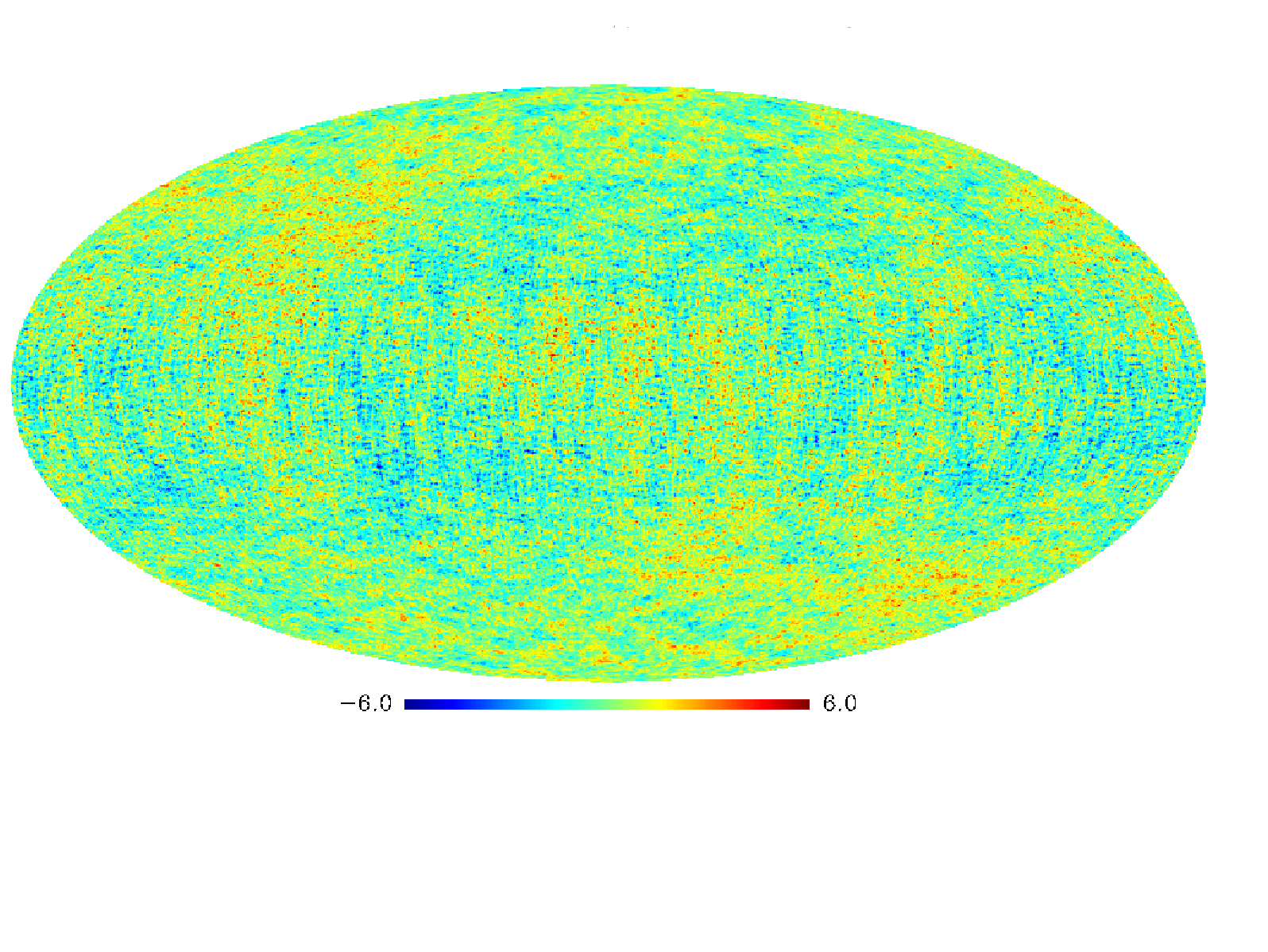}
} \vskip 0.25 truein \caption{Noise-free simulation of a model
with scale-invariant quadrupole asymmetry in the primordial power
with $g_{LM} = 0.1 \delta_{L2}\delta_{M0}$. The isotropic
component of the temperature map is shown on the left, and the
anisotropic component on the right. The colour scales are in
$\protect\mu \mathrm{K}$. } \label{mapT}
\end{figure}

\begin{figure}[tbp]
\centerline{\includegraphics[bb=15 37 743
440,width=3.4in]{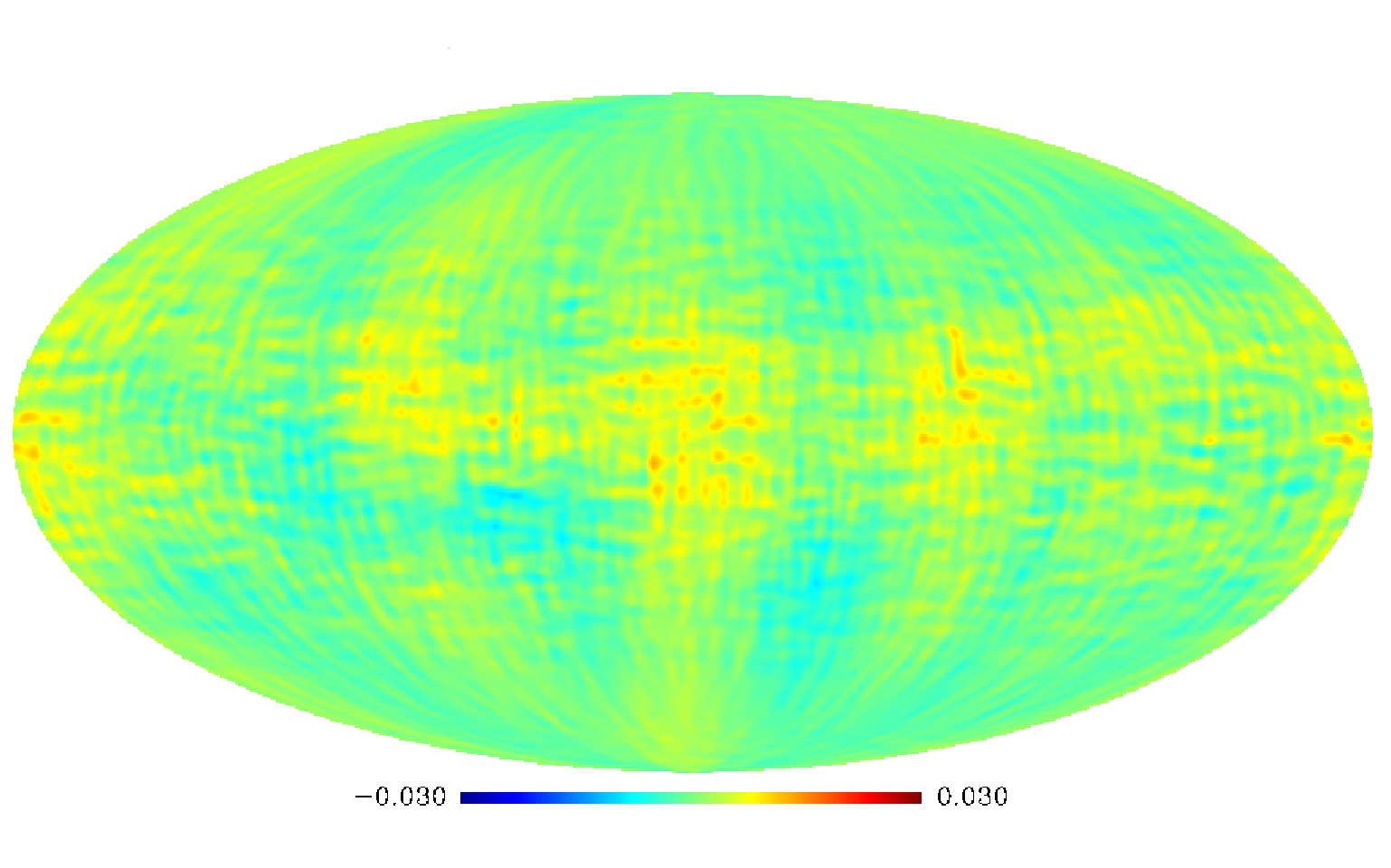}
\includegraphics[bb=15 37 743 440,width=3.4in]{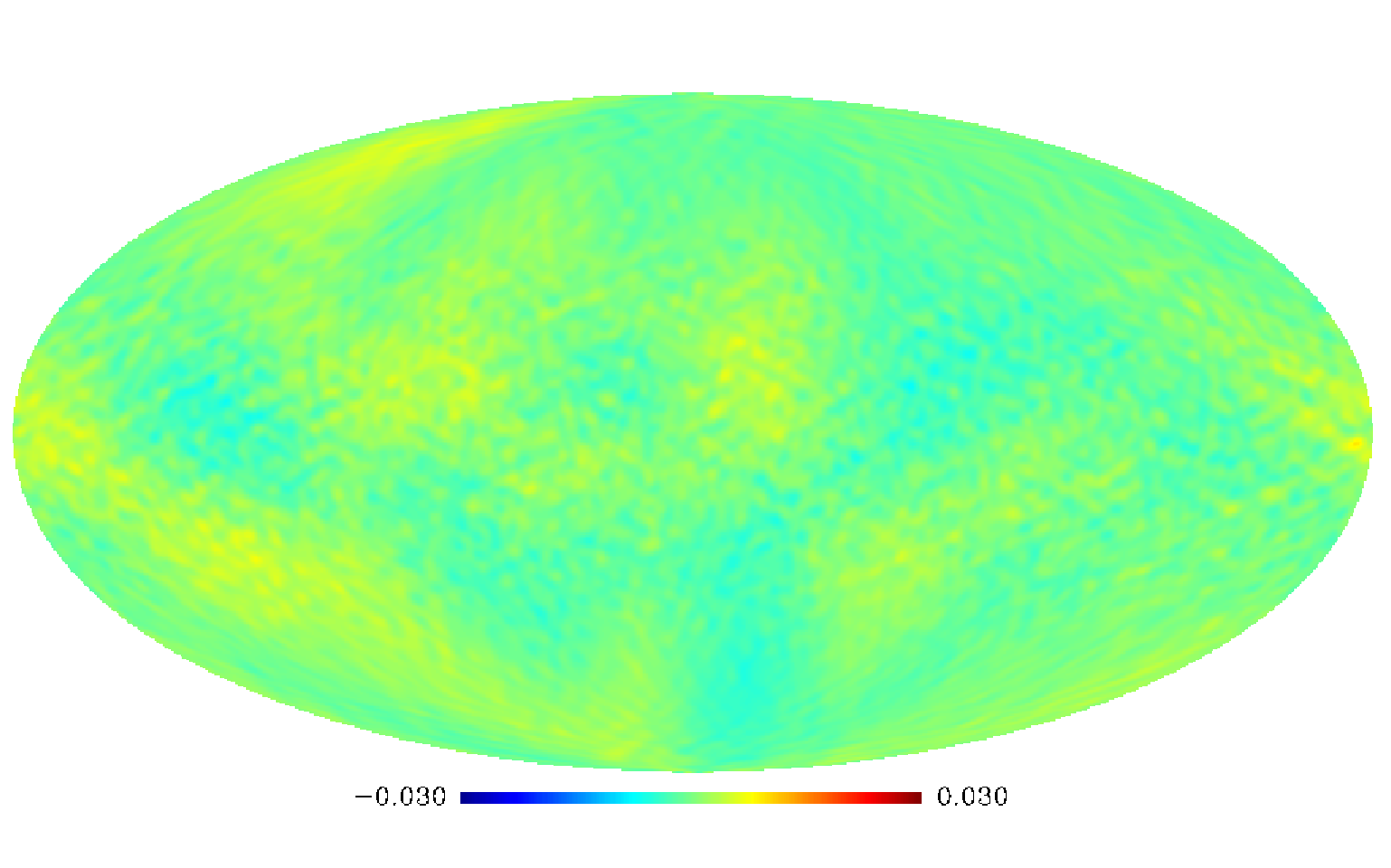}} \vskip %
0.25 truein \caption{Anisotropic components of the $Q$ (left) and
$U$ (right) polarization maps in a noise-free simulation of the
model in Fig.~\ref{mapT}. The maps have been smoothed here with a
Gaussian beam of FWHM $3^\circ$ to enhance the imprint of the
preferred axis in the $Q$ map. The colour scales are in
$\protect\mu \mathrm{K}$. } \label{mapQU}
\end{figure}

To give a feel for the nature of the anisotropy signal, Figs.
\ref{mapT} and \ref{mapQU} show simulated sky maps for a
noise-free realisation of a scale-invariant ($q=0$)
quadrupole-modulation model with $g_{20}=0.1$ ($g_{2M}=0$, $M\neq
0$). These maps have been generated using the prescription described in \cite%
{Hanson09}, generalized to polarization. This uses an approximate
square root of the anisotropic covariance matrix, linear in the
$g_{LM}$, to simulate maps as the sum of a statistically isotropic
part and an anisotropic part. The isotropic component of a
noise-free temperature map is shown on the left in Fig.~\ref{mapT}
and the anisotropic component, clearly showing a preferred
direction along the polar axis, is shown on the right.

The anisotropic contribution to the Stokes $Q$ and $U$
polarization maps is shown in Fig.~\ref{mapQU}. Here, we have
smoothed the polarization maps with a Gaussian of FWHM 3 deg.\ to
enhance the visual impact of the statistical anisotropy in the $Q$
Stokes map.  Because of the quadrupole asymmetry in the primordial
power, modes of the primordial perturbation with wavevectors along
the polar axis tend to have their amplitude enhanced. For such
modes, the polarization generated by Thomson scattering is pure
$Q$ in the polar basis and varies in amplitude with the polar
angle as $\sin^2\theta$~\cite{Hu97}. The dominant effect of the
statistical anisotropy is therefore observed in the $Q$ Stokes
parameter and is concentrated toward the equatorial plane.

\begin{figure}[tbp]
\centerline{\includegraphics[bb=0 0 729
480,width=3.0in]{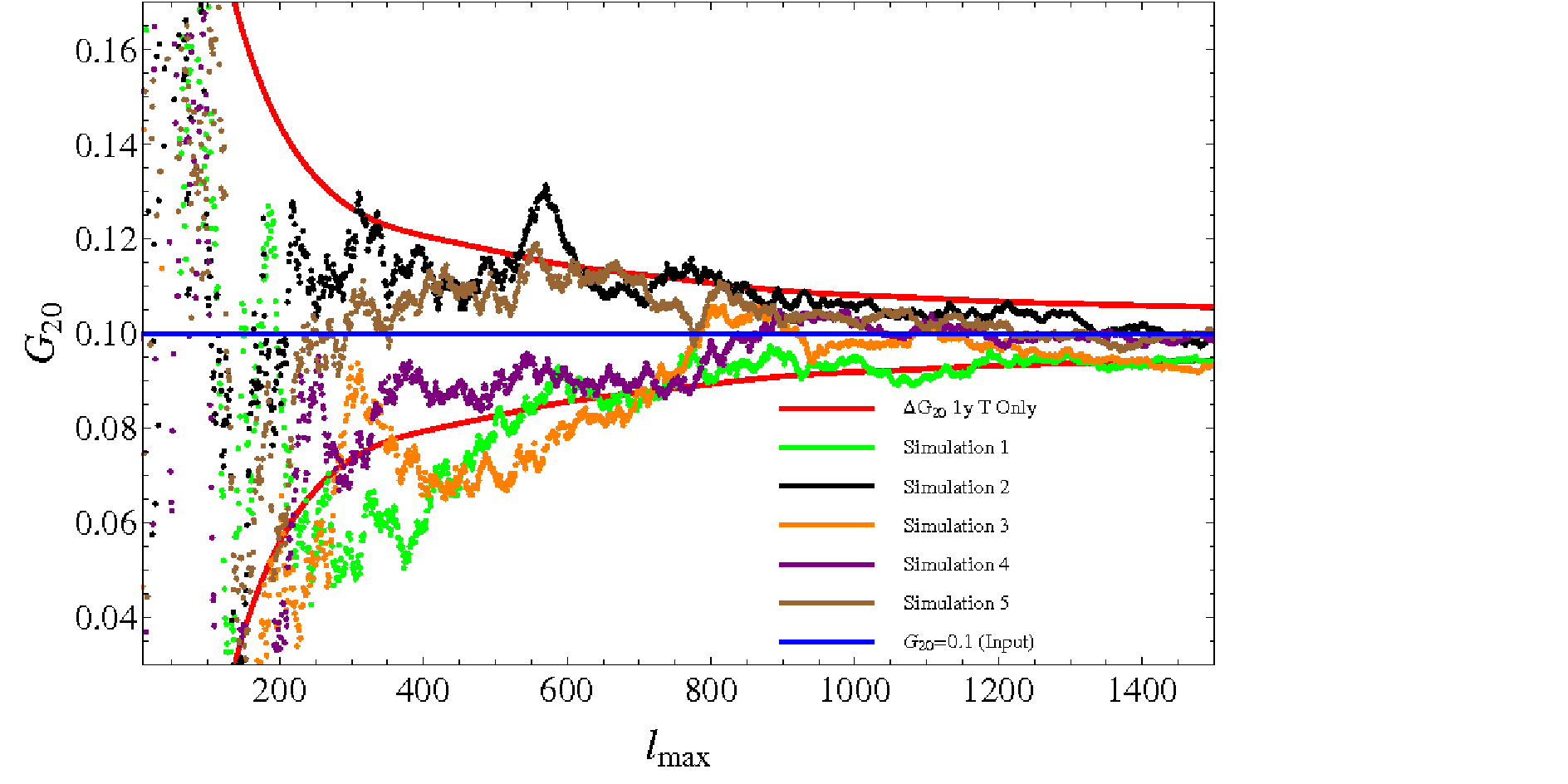}
\includegraphics[bb=0 0 611
396,width=3.1in]{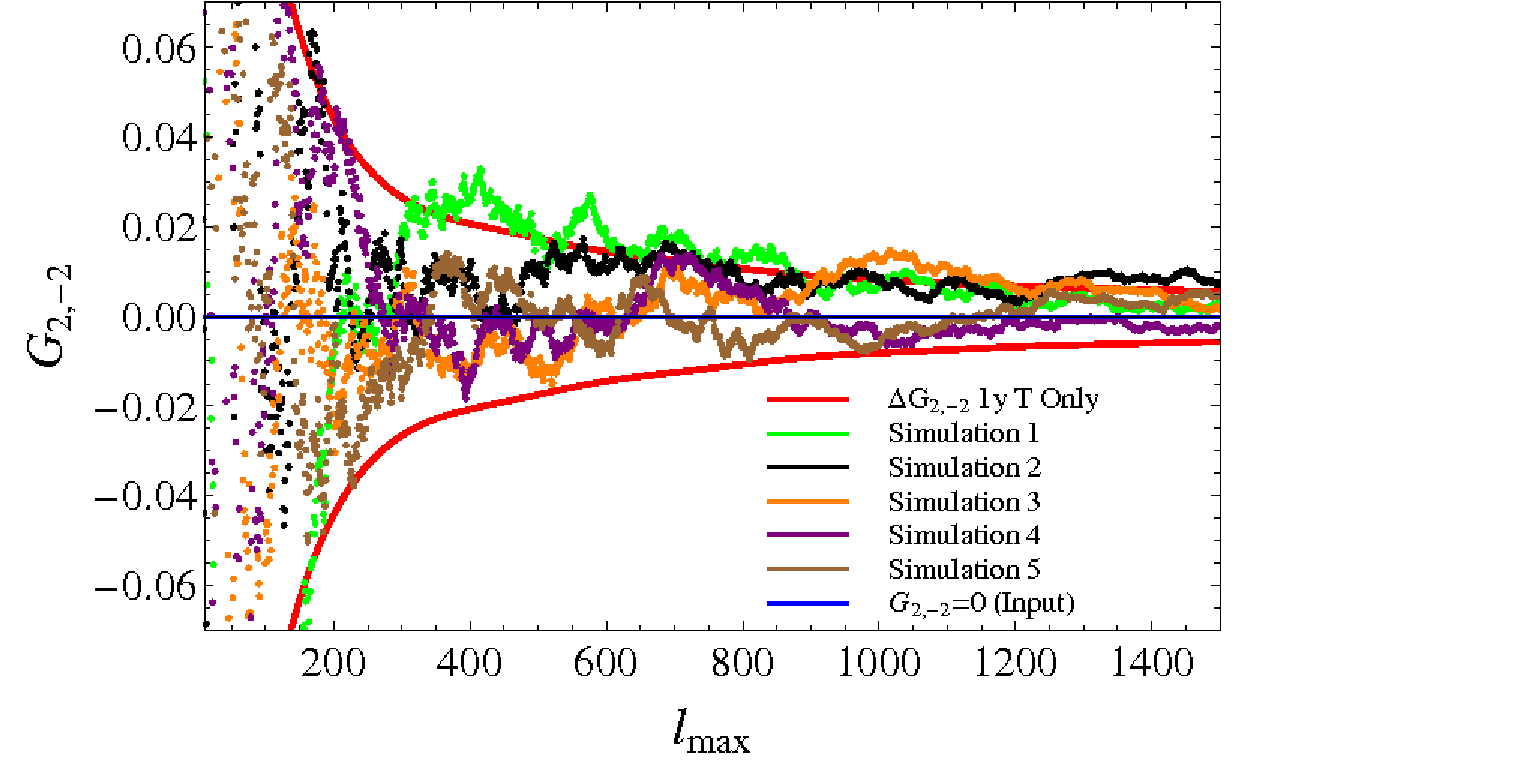}} \centerline{\includegraphics[bb=0 0
800 524,width=3.0in]{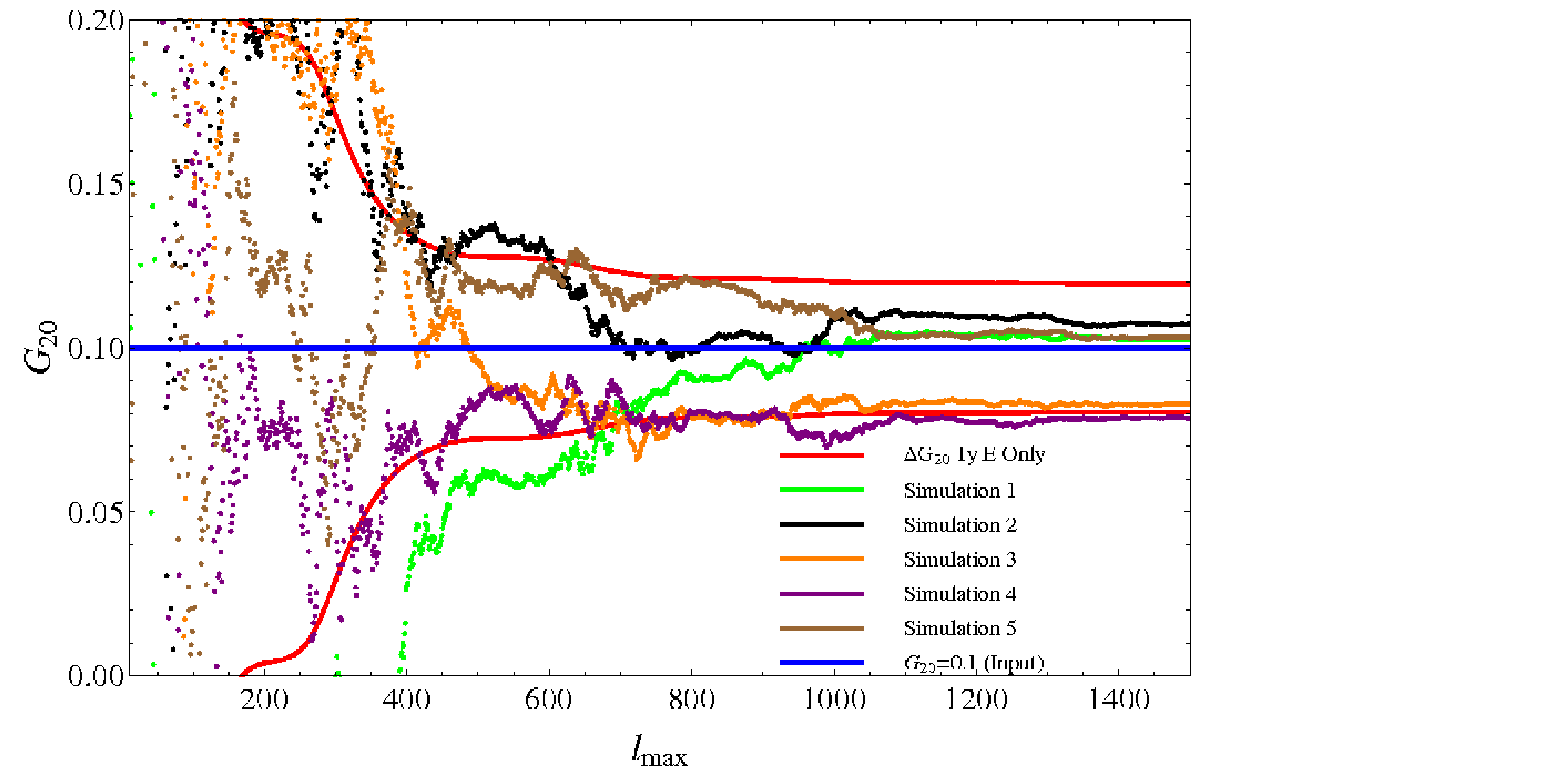}
\includegraphics[bb=0 0 730 475,width=3.1in]{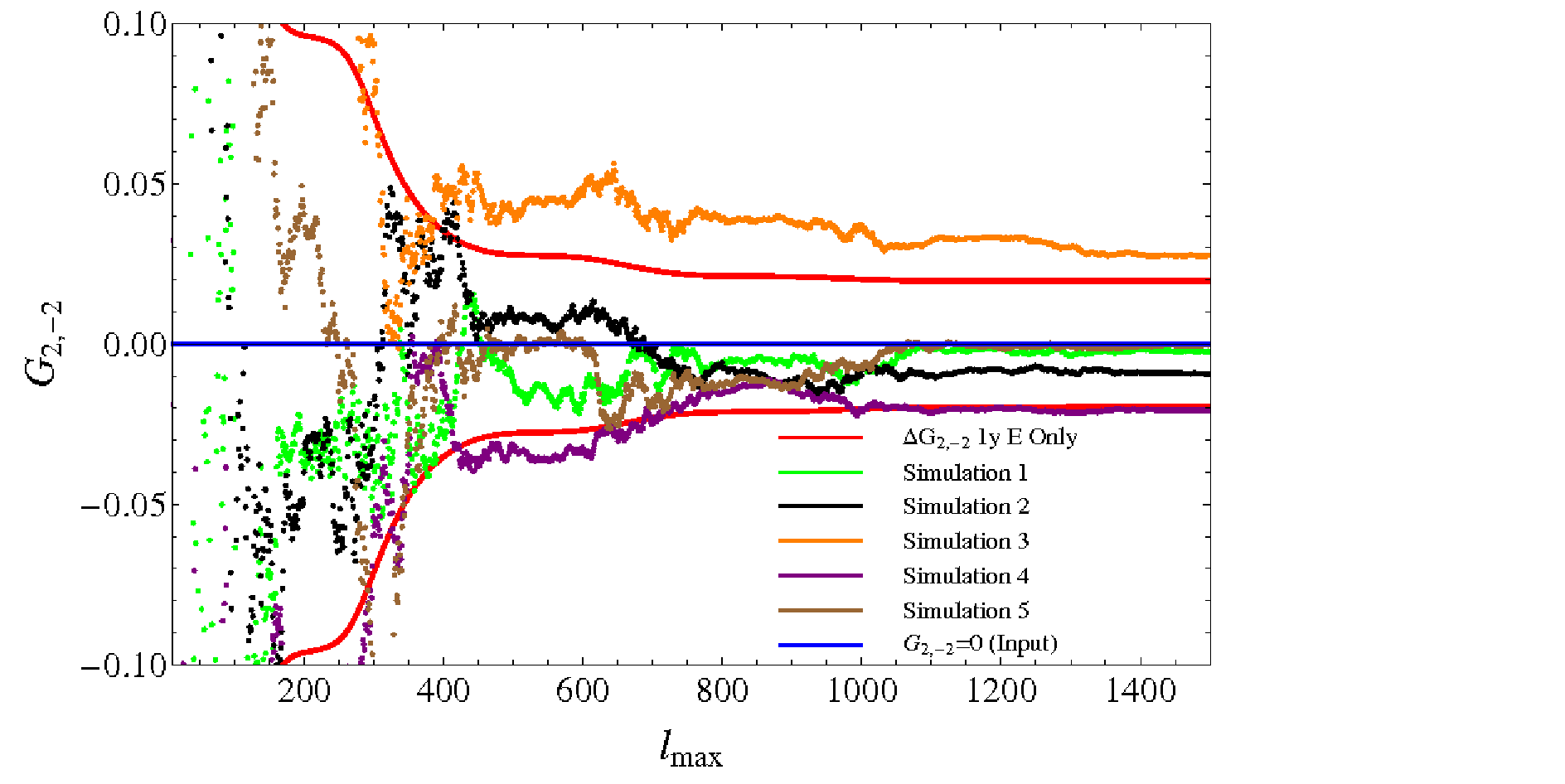}}
\centerline{\includegraphics[bb=0 0 722
472,width=3.0in]{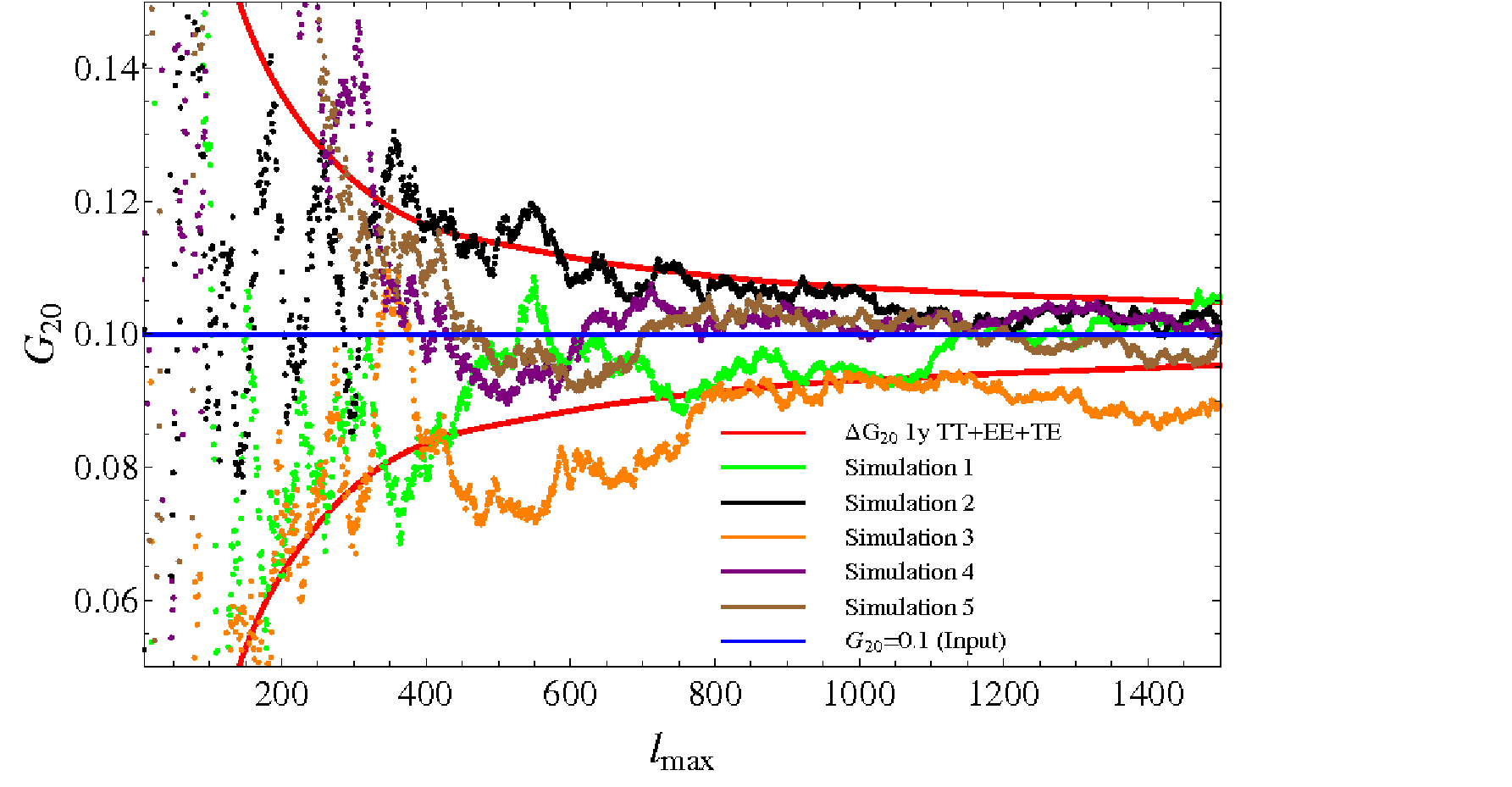}
\includegraphics[bb=0 0 751 485,width=3.1in]{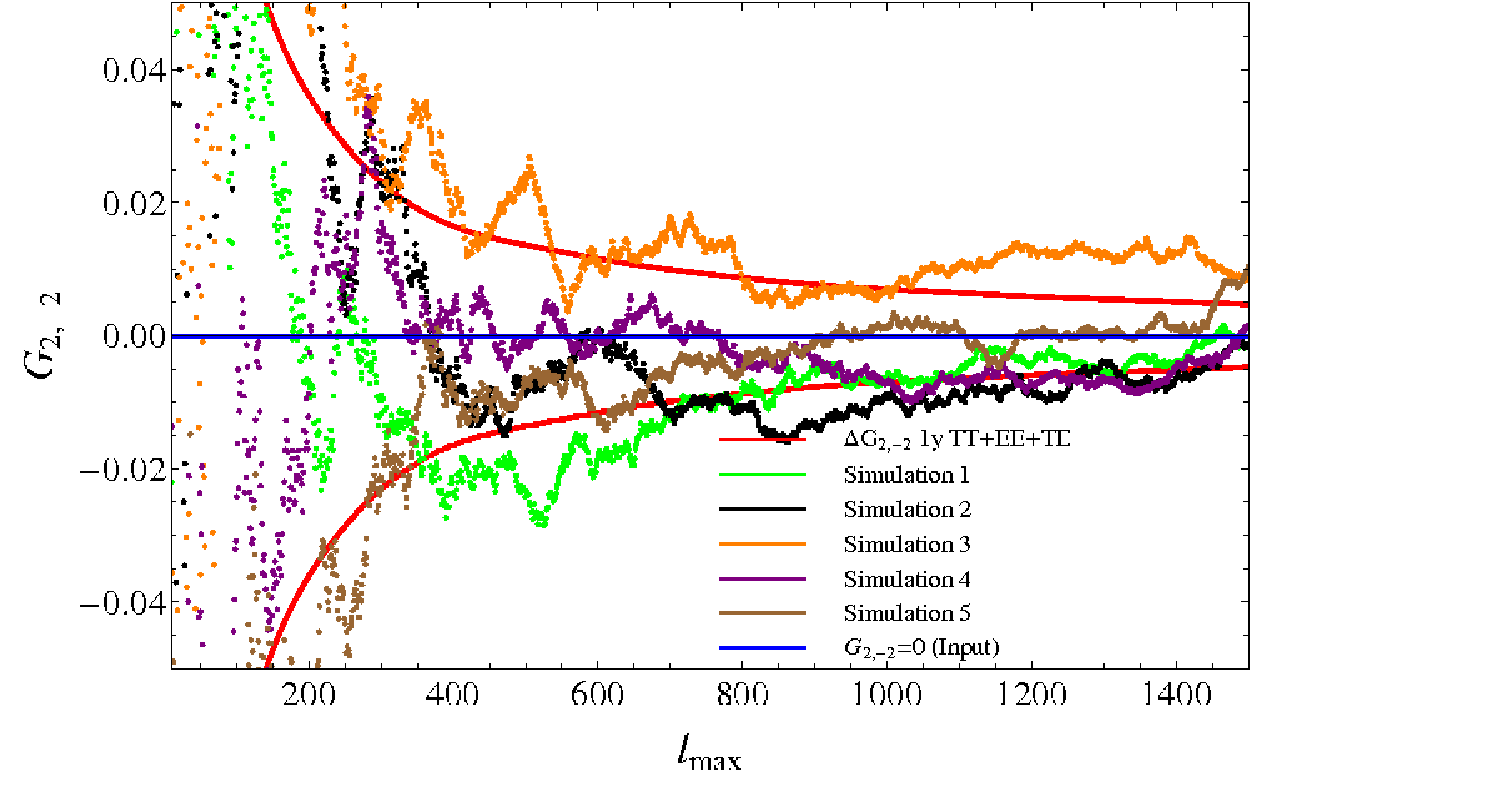}}
\caption{Estimates of the $G_{20}$ (left) and $G_{2\,-2}$ (right)
anisotropy parameters (shown with points) and their (one-sigma)
Fisher errors ([Red] solid lines) as a function of
$l_{\mathrm{max}}$ from five simulations of the model in
Fig.~\ref{mapT} for one year of Planck data. The input parameters
$G_{20}=0.1$ and $G_{2\,-2}$ are shown with horizontal [Blue]
solid lines. From top to bottom we analyse temperature only,
$E$-mode polarization only and temperature plus polarization.}
\label{g2ms}
\end{figure}

To illustrate the machinery summarised in Section
\ref{section-estimator}, we have generated five simulations of the
scale-invariant quadrupole-modulation model with $g_{2M}=0.1\delta
_{0M}$ and added instrumental noise appropriate to one year of
observation with the Planck $143\;\mathrm{GHz}$ channel. We then
estimate $g_{2M}$ via Eq.~(\ref{eq:estimator}) as a function of
$l_{\mathrm{max}}$, the maximum multipole retained in the
analysis. Since the only nonzero coefficient in these simulations
is $g_{20}$, and the survey is assumed isotropic, the recovered
estimates $\hat{g}_{2M}$ are statistically equivalent for $M\neq
0$ and so we show results only for the two (real) components
$\hat{G}_{20}$ and $\hat{G}_{2-2}$ in Fig.~\ref{g2ms}. We analyse
temperature alone (top panels), $E$-mode polarization alone
(middle), and both jointly (bottom).

With temperature alone, the errors on $G_{2M}$ decrease
approximately as $1/l _{\mathrm{max}}$ over the range plotted in
Fig.~\ref{g2ms} ($l \leq 1500$)  reaching $0.005$ by
$l_{\mathrm{max}}$ (in agreement with the minimum-variance
estimators of~\cite{Pullen07}). This behaviour follows from simple
mode-counting since the temperature maps
are signal-dominated over this multipole range.\footnote{%
For modes that are signal-dominated, the scale dependence of the
trace term in the Fisher matrix, Eq.~(\ref{FM2}), is weak.
Treating the trace as constant gives Fisher information varying as
$l_{\mathrm{max}}^2$ which is proportional to the number of modes
retained in the analysis.} However, the polarization maps are
noise-dominated over much of this multipole range and so the
errors approach constant values for
$l_{\mathrm{max}}\lower.5ex%
\hbox{\gtsima}600$. Nevertheless, the Planck polarization maps
alone can provide (almost) independent constraints on an
anisotropic modulation to the temperature maps. For two sky
surveys, the errors on the $g_{2M}$ from polarization are four
times worse than in temperature. Consistency between temperature
and polarization constraints would provide an important test of
systematic effects should Planck show any evidence of an
anisotropic power spectrum.
\subsection{Constraints on scale-dependence}
\begin{figure}[tbp]
\centerline{\includegraphics[bb=0 0 512
337,width=2.2in]{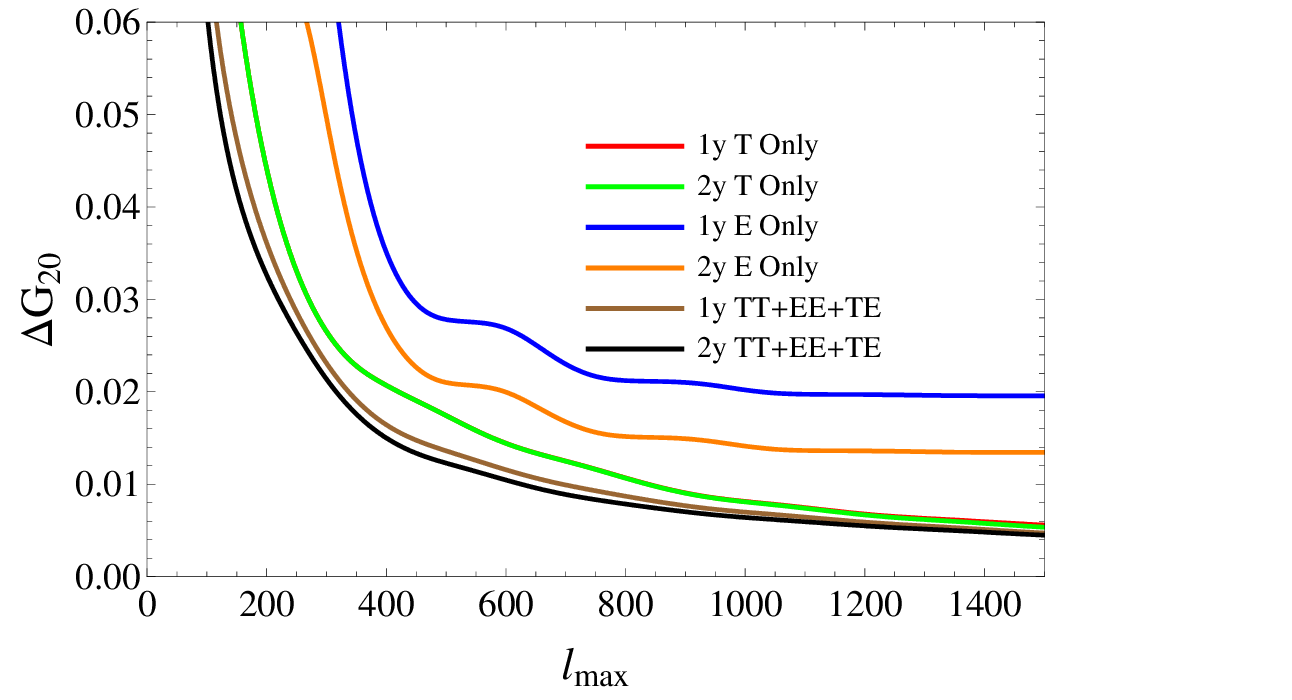}
\includegraphics[bb=0 0 535
350,width=2.2in]{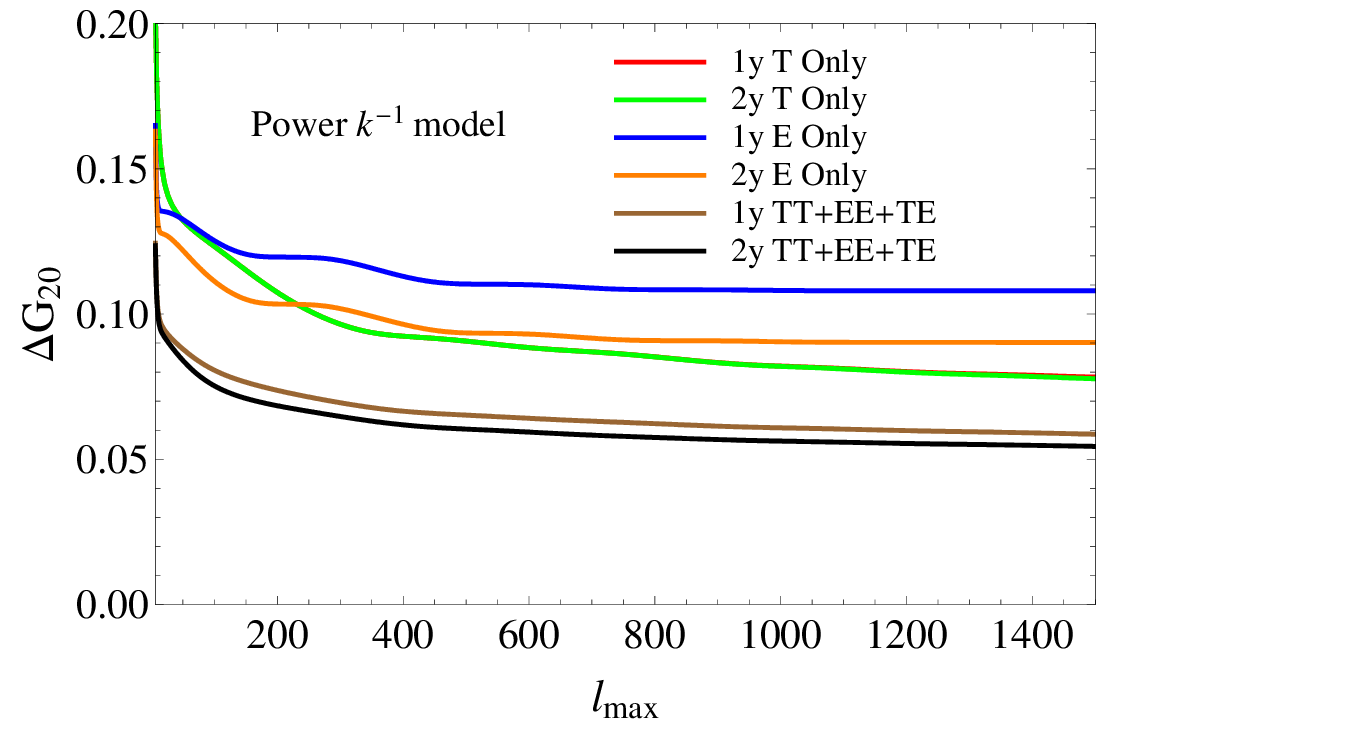}
\includegraphics[bb=0 0 464
300,width=2.25in]{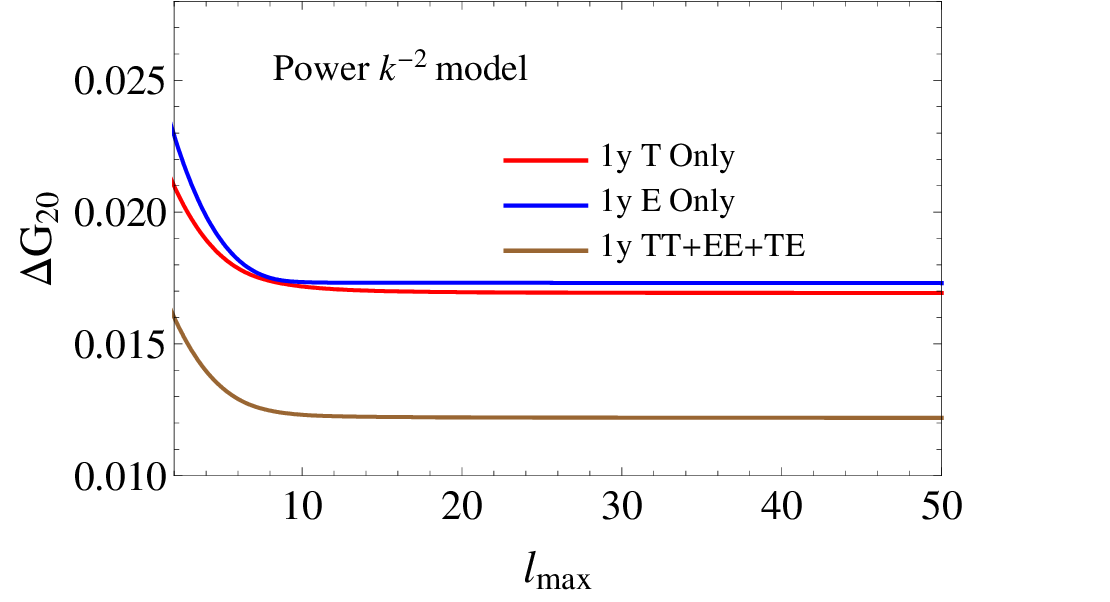}} \caption{Fisher errors for
$G_{20}$ from temperature, $E$-mode polarization, and temperature
plus polarization in models with power-asymmetry spectral indices
$q=0$ (left), $q=1$ (middle) and $q=2$ (right). For $q=0$ and
$q=1$ we show results for one and two years of observations; for
$q=2$ we show only the one-year errors since they improve very
little with further observing time. Note that the one- and
two-year errors from temperature alone are indistinguishable when
$q\geq 0$. \label{error-comparsion}}
\end{figure}

In Fig.~\ref{error-comparsion}, we compare the Fisher errors on
the amplitude of the modulation for a scale-invariant model
($q=0$) and two models with scale-dependence ($q=1$ and $q=2$).
For larger $q$, the asymmetry in the variance of the Fourier modes
is confined to larger scales and so relatively more of the
constraining power derives from low-$l$ multipole moments. The
low-$l$ modes of polarization are enhanced by scattering at
reionization~\cite{Zaldarriaga97} and are expected to be
signal-dominated in the one-year Planck data. The polarization
constraints on the $g_{2M}$ therefore become more comparable to
those from the temperature as $q$ increases and the improvement
from observing for longer in polarization lessens.

If the amplitude of any primordial power asymmetry is high enough,
it might be possible to constrain the scale-dependence of the
asymmetry with Planck. To forecast constraints on the spectral
index of the power asymmetry, $q$, we extend the Fisher matrix
analysis of Sec.~\ref{section-estimator} to include $q$ as a
parameter. We must now evaluate the Fisher matrix at nonzero
$g_{LM}$ but we assume that the asymmetry is still small enough
that we can neglect asymmetry in the
$(\mathbf{C}^{\mathrm{tot}})^{-1}$ terms. The $F_{LM,L'M'}$ is
then unchanged from Eq.~(\ref{FM2}) but the additional elements
are
\begin{eqnarray}
F_{LM,q} &=& \frac{1}{2} \sum_{l_1 m_1}\sum_{l_2
m_2}\mathrm{Tr}\left[(\mathbf{C}_{l_1}^{\mathrm{tot}})^{-1}
\frac{\partial \mathbf{C}_{l_1 m_1, l_2 m_2}}{\partial g_{LM}^*}
(\mathbf{C}_{l_2}^{\mathrm{tot}})^{-1} \frac{\partial
\mathbf{C}_{l_2 m_2, l_1 m_1}}{\partial q}
\right]  \nonumber \\
&=&\sum_{l_{1}l_{2}}\left[ \text{Tr}\left(
\tilde{\mathbf{C}}_{l_{1}l_{2}}
(\mathbf{C}_{l_{2}}^{\mathrm{tot}})^{-1}\partial_q
\tilde{\mathbf{C}}_{l_{2}l_{1}}
(\mathbf{C}_{l_{1}}^{\mathrm{tot}})^{-1}\right) g_{LM} \frac{%
(2l_{1}+1)(2l_{2}+1)}{8\pi }\left(
\begin{array}{ccc}
l_{1} & l_{2} & L \\   \label{fglmq}
0 & 0 & 0%
\end{array}%
\right)^{2}\right] , \\
F_{q,q} &=& \frac{1}{2} \sum_{l_1 m_1}\sum_{l_2
m_2}\mathrm{Tr}\left[(\mathbf{C}_{l_1}^{\mathrm{tot}})^{-1}
\frac{\partial \mathbf{C}_{l_1 m_1, l_2 m_2}}{\partial q}
(\mathbf{C}_{l_2}^{\mathrm{tot}})^{-1} \frac{\partial
\mathbf{C}_{l_2 m_2, l_1 m_1}}{\partial q}
\right]  \nonumber \\
&=&\sum_{l_{1}l_{2}} \left [  \text{Tr}\left(\partial_q
\tilde{\mathbf{C}}_{l_{1}l_{2}}
(\mathbf{C}_{l_{2}}^{\mathrm{tot}})^{-1}\partial_q
\tilde{\mathbf{C}}_{l_{2}l_{1}}
(\mathbf{C}_{l_{1}}^{\mathrm{tot}})^{-1}\right) \frac{%
(2l_{1}+1)(2l_{2}+1)}{8\pi }  \right. \nonumber \\
& &  \left.  \qquad \qquad \times \sum_{LM} |g_{LM}|^2
\left(\begin{array}{ccc} l_{1} & l_{2} & L \cr 0 & 0 & 0
\end{array}%
\right )^{2}  \right ]   . \label{fqq}
\end{eqnarray}
Note that these elements vanish if $g_{LM}=0$. From the scaling of
the Fisher matrix elements with $g_{LM}$, we expect the error on
$q$ to scale as the inverse of the amplitude of the asymmetry.


As an example, we consider a fiducial model with scale-invariant
quadrupole asymmetry with $G_{2M} = 0.03 \delta_{M0}$. Such a
model is compatible with the constraints on quadrupole asymmetry
from the beam-corrected analysis of WMAP data in
Ref.~\cite{Hanson10}, but the amplitude should be detectable with
Planck at the $6\sigma$ level (fixing $q=0$). Forecasts for the
(marginalised) errors on $q$ for this model are given in
Table~\ref{tab1}.  With temperature and polarization, Planck
should constrain the spectral index to a $1\sigma$ accuracy of
$\Delta q \sim 0.3$. The marginalised errors on the $g_{LM}$ are
similar to the case in which $q$ is fixed.

\begin{table*}[tbp]
\begin{centering}
\begin{tabular}{c|c|c|c}\hline
$\Delta q$ ($1\sigma$) & $T$ only & $E$ only & $T$ and $E$
\\\hline\hline
One year & $0.399$ & $1.300$ & $0.322$ \\\hline Two years &
$0.389$ & $0.878$ & $0.299$\\ \hline
\end{tabular}%
\caption{Fisher errors on the scale-dependence of the power
asymmetry assuming a scale-invariant quadrupole asymmetry with
$g_{2M} = 0.03 \delta_{M0}$.} \label{tab1}
\end{centering}
\end{table*}

\subsection{Axisymmetric models}
\label{pre-axis} The preceding analysis makes no assumptions about
axisymmetry of the primordial power asymmetry. However, if we have
good reason to expect axisymmetry, so the model is described by a
preferred axis $\hat{\mathbf{m}}$ and the $M=0$ multipoles
$\{g_{*L}\}$ in a frame with the polar axis along
$\hat{\mathbf{m}}$, we can constrain these parameters by
post-processing our estimates $\hat{g}_{LM}$. We illustrate how
this works,  assuming a fixed scale dependence for the primordial
asymmetry.

For models with a nearly scale-invariant anisotropy spectrum, many
small-scale modes contribute to the $\hat{g}_{LM}$, so we might
expect the statistics of the $\hat{g}_{LM}$ to be approximately
Gaussian.  Expressing $\hat{\mathbf{m}}$ in terms of its azimuthal
angle $\alpha$ and polar angle $\beta$,
$\hat{\mathbf{m}}=D(\alpha,\beta,0)\hat{\mathbf{z}}$ (i.e.\ a
rotation of the $\hat{\mathbf{z}}$ direction through Euler angles
$\alpha$ and $\beta$), in the Gaussian approximation we can write
$\mathrm{Pr}(\{\hat{g}_{LM}\}|\hat{\mathbf{m}}, \{g_{*L}\})
\propto \exp(-\chi^2/2)$ where
\begin{equation}
\chi^2  = \sum_{LM} \sum_{L'M'} (\hat{g}_{LM}^* -
\tilde{g}_{LM}^*)F_{LM,L'M'} (\hat{g}_{L'M'}-\tilde{g}_{L'M'}) .
\label{eq:chisq}
\end{equation}
Here,
\begin{equation}
\tilde{g}_{LM} \equiv  D_{M0}^L(\alpha,\beta,0)g_{*L}
\label{eq:rotn}
\end{equation}
are the multipoles of the primordial asymmetry rotated from their
preferred frame. ($D^L_{MM'}(\alpha,\beta,\gamma)$ are the Wigner
rotation matrices.) If we now assign a uniform prior on the
direction $\hat{\mathbf{m}}$ [so that $\mathrm{Pr}(\alpha,\beta)
d\alpha d\beta = (4\pi)^{-1} d \alpha d\cos\beta = (4\pi)^{-1} d
\hat{\mathbf{m}}$] and a flat prior on the $g_{*L}$, Bayes'
theorem gives for the posterior
\begin{eqnarray}
\mathrm{Pr}(\hat{\mathbf{m}},\{g_{*L}\}|\{\hat{g}_{LM}\})
d\hat{\mathbf{m}} &=&
\mathrm{Pr}(\alpha,\beta,\{g_{*L}\}|\{\hat{g}_{LM}\})d\alpha d\beta \nonumber \\
&\propto& e^{-\chi^2/2} d\hat{\mathbf{m}} .
\end{eqnarray}

For an isotropic survey (and assumed weak anisotropy), the Fisher
matrix is isotropic and we can write $F_{LM,L'M'} =
\delta_{LL'}\delta_{MM'}/\sigma_L^2$. Substituting
Eq.~(\ref{eq:rotn}) into Eq.~(\ref{eq:chisq}) and using the
addition theorem for the rotation matrices, we have
\begin{equation}
\chi^2 = \sum_{L} \frac{1}{\sigma_L^2} \left(g_{*L}^{2}-2g_{*L}
\sum_{M}\Re \left[\hat{g}_{LM}D_{M0}^{L\ast }(\alpha ,\beta ,0)
\right]+\sum_{M}|\hat{g}_{LM}|^2\right) . \label{like_glm2}
\end{equation}%
Noting that $D_{M0}^{L\ast }(\alpha ,\beta ,0)=\sqrt{4\pi
/(2L+1)}Y_{LM}(\hat{\mathbf{m}} )$, if we define
\begin{equation}
\hat{g}_L(\hat{\mathbf{m}}) \equiv \sum_{M} \hat{g}_{LM}
Y_{LM}(\hat{\mathbf{m}}) ,
\end{equation}
(i.e.\ a map of the estimated $\hat{g}_{LM}$ at multipole $L$), we
can write the posterior as
\begin{eqnarray}
\mathrm{Pr}(\hat{\mathbf{m}},\{g_{*L}\}|\{\hat{g}_{LM}\})
d\hat{\mathbf{m}} &\propto & \prod_L \exp \left( -\frac{2\pi
}{(2L+1)\sigma _{L}^{2}}\left[ g_{*L}Y_{L0}(\hat{\mathbf{z}})-
\hat{g}_L(\hat{\mathbf{m}})\right]^2 \right) \nonumber \\
&&\mbox{} \qquad\qquad \times \exp \left( \frac{2\pi
}{(2L+1)\sigma _{L}^{2}}\hat{g}_L^{2}(\hat{\mathbf{m}})\right)
d\hat{\mathbf{m}} .
\end{eqnarray}
The marginal distribution for the direction of the axis is given
by integrating over $g_{*L}$:
\begin{equation}
\mathrm{Pr}(\hat{\mathbf{m}}|\{\hat{g}_{LM}\}) d \hat{\mathbf{m}}
\propto \prod_L \exp \left( \frac{2\pi }{(2L+1)\sigma
_{L}^{2}}\hat{g}_L^{2}(\hat{\mathbf{m}})\right) d\hat{\mathbf{m}}
,
\end{equation}
so that contours of constant density are given by the contours of
$\sum_L \hat{g}_L^{2}(\hat{\mathbf{m}})/[(2L+1)\sigma_L^2]$. For
the simple case of a  primordial power asymmetry at a single
multipole $L$, e.g.\ a quadrupole asymmetry, $\ln
\mathrm{Pr}(\hat{\mathbf{m}}|\{\hat{g}_{LM}\})$ is proportional to
the square of the map of the reconstructed multipoles.

\begin{figure}[tbp]
\centerline{%
\includegraphics[bb=0 0 483 463, width=2.3in]{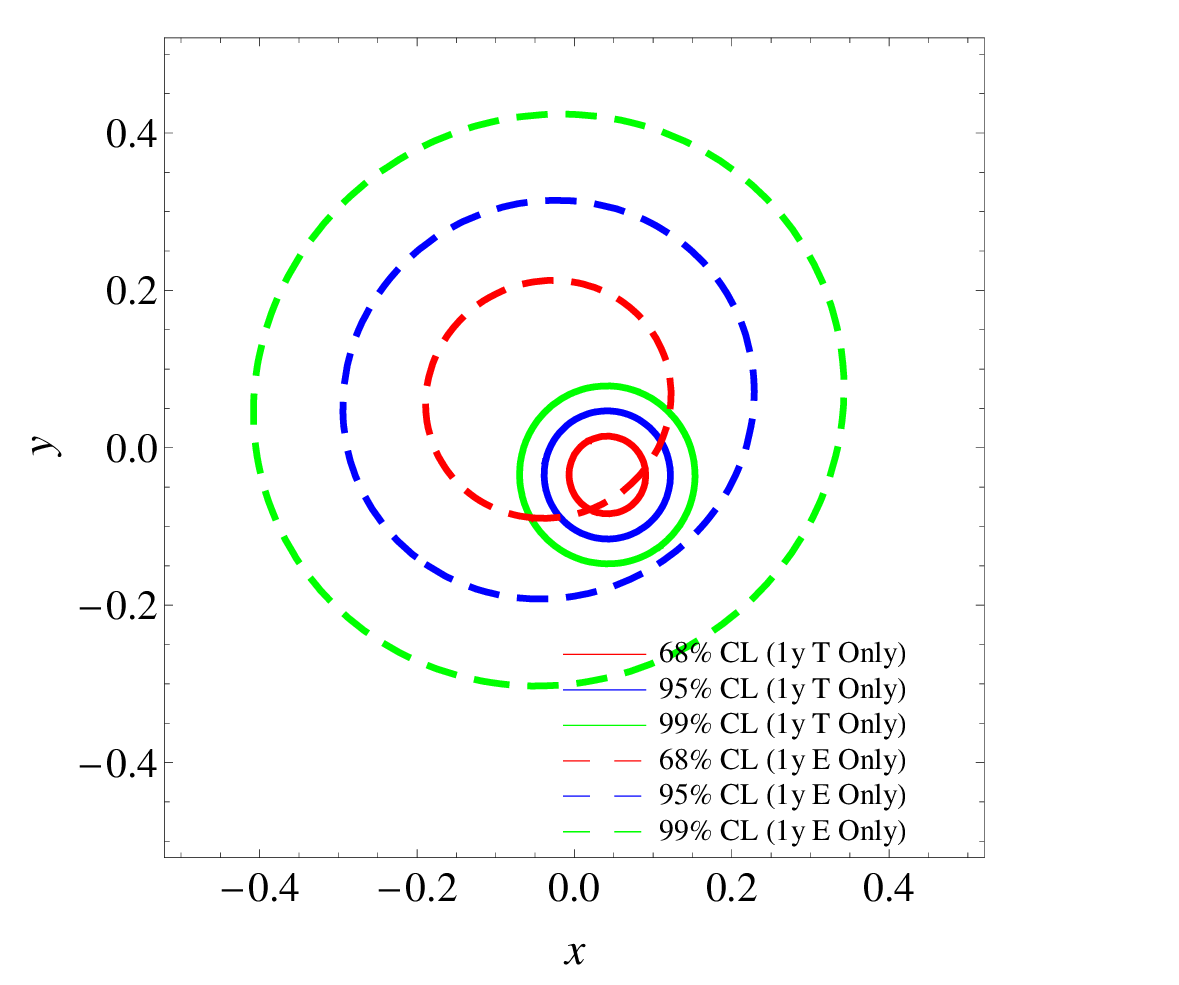}
\includegraphics[bb=0 0 498 337,width=3.2in]{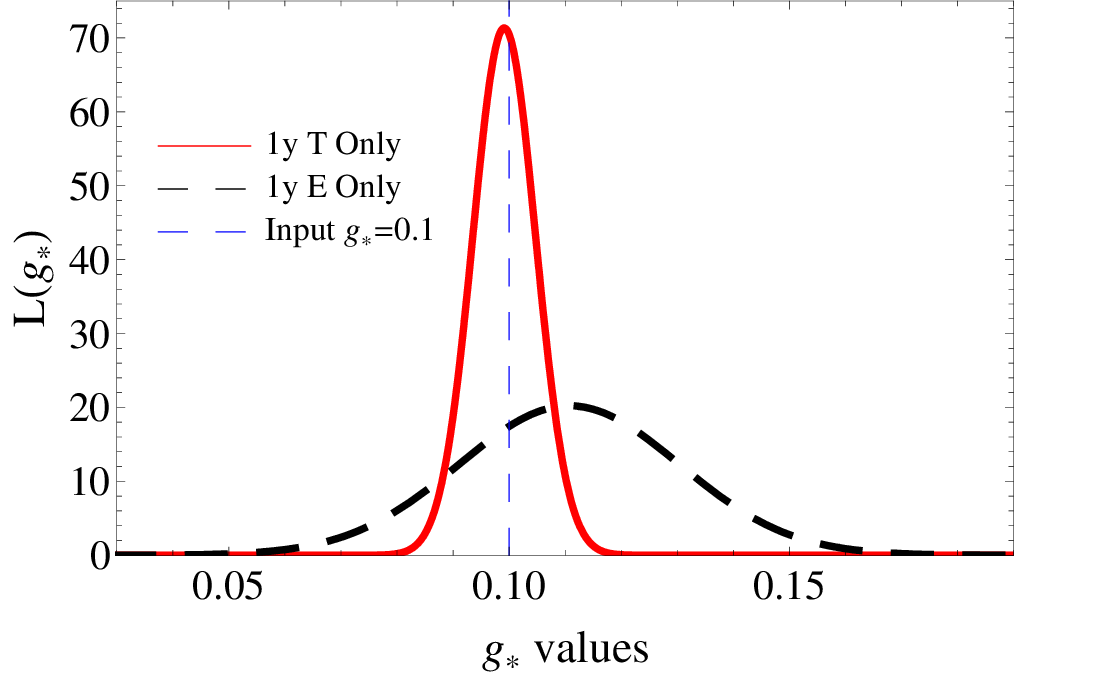}}
 \caption{Marginal distributions for the direction (left) and
amplitude ($g_{*2}$, right) from a simulation of the nominal
(one-year) Planck survey for a model with an axisymmetric
quadrupole asymmetry aligned with the polar axis ($g_{LM}=0.1
\delta_{L2} \delta_{M0}$). We parameterise the direction with the
equal-area projection $x=2\sin(\beta/2)\cos(\alpha)$ and
$y=2\sin(\beta/2)\sin(\alpha)$ and show in solid lines the 68\%
[Red], 95\% middle [Blue] and 99\% outer [Green] contours from the
temperature alone; dashed-line contours are from the $E$-mode
polarization alone. For the amplitude (right), we plot the
marginal distributions from $T$ alone (solid line) and $E$ alone
(dashed line).} \label{dis-g}
\end{figure}

To illustrate these ideas, in Fig.~\ref{dis-g} we plot the
marginal distributions for the direction and amplitude of the
model analysed in Sec.~\ref{subsec:amp_cons} using one of the
simulations shown in Fig.~\ref{g2ms}.  We parameterise the
direction by the Cartesian components of the equal-area projection
$x=2\sin(\beta/2)\cos(\alpha)$ and $y=2\sin(\beta/2)\sin(\alpha)$.
With one year of simulated Planck temperature data, the
constraints on the direction of the axis are $\Delta x \approx
\Delta y = 0.032$ (i.e.\ $1.8^\circ$) at 68\% confidence and the
$x$ and $y$ components are almost uncorrelated as expected from
isotropy. The amplitude  $g_{*2}=0.1 \pm 0.005$ (68\% confidence)
with the error being very nearly $\sigma_2$. With polarization
alone, these constraints weaken to $\Delta x \approx \Delta y =
0.10$ (i.e.\ $5.7^\circ$) and $g_{*2}=0.11 \pm 0.02$. The maximum
of the posterior in this simulation is at $x=0.042$, $y=-0.035$
and $g_{*2}=0.099$ ($T$ only) and $x=-0.033$, $y=0.062$ and
$g_{*2}=0.11$ ($E$ only) and the $\chi^2$ at these values are
$0.223$ and $0.236$ respectively, with $5-3=2$ degrees of freedom.

\section{Conclusions}
\label{section-conclusion} As summarized in Sec.~\ref{sec:intro},
cosmological perturbations are usually assumed to satisfy
statistical isotropy, as expected in simple models of inflation.
Yet there is some tentative observational evidence from the CMB
suggesting possible departures from statistical isotropy. Here we
have shown that Planck should be able to set strong constraints on
small departures from statistical anisotropy  and, under
favourable circumstance, could set constraints on their scale
dependence and any preferred direction.

In this paper we have developed and applied quadratic estimators
to test for an asymmetry in the primordial power spectrum from
temperature and polarization measurements of the CMB. Our
estimators are optimal in the limit of isotropic primordial power.
We have tested our methods against simulations that include a
quadrupole power asymmetry.

We have analysed the ability of the Planck mission to constrain
models with quadrupole power asymmetry using temperature and
polarization data.  Using temperature data alone, Planck should be
able to constrain each multipole $g_{2M}$ of a scale-invariant
quadrupole anisotropy at the $0.01$ level ($2\sigma$), well below
the current constraints derived from WMAP ($|g_{2M}| < 0.07$ from
Ref.~\cite{Hanson10}). Using polarization data alone from an
extended Planck mission (four sky surveys) such an anisotropy can
be constrained to an accuracy only about three times worse than
from the temperature. This offers the possibility of a consistency
check on the existence on any observed departure from statistical
isotropy.

If the amplitude of a power asymmetry is large enough, it may be
possible to constrain its scale dependence. We have estimated the
Fisher errors when additionally constraining a free spectral index
describing a power law modulation, $g_{LM}(k) = g_{LM} (k_0/k)^q$.
For a scale-free quadrupole modulation with an amplitude of 1\%
(i.e.\ $g_{20}/\sqrt{4\pi} = 0.01$ in an axi-symmetric model), we
find that an extended Planck mission can constrain the spectral
index to a $1\sigma$ accuracy of $\Delta q \sim 0.3$.

Finally, we have considered the constraints on a preferred
direction in models with a purely axisymmetric modulation of the
primordial power spectrum. In a  scale-free model with a $1\%$
quadrupole modulation, the direction of the preferred axis can be
determined from Planck data to a precision of about $2^\circ$
using temperature observations alone and to about $6^\circ$ using
polarization data alone.

The quadratic estimators developed here for isotropic surveys can,
in principle, be straightforwardly extended to deal with
real-world effects such as anisotropic noise and Galactic masks.
However, the calculation of the $\bar{a}^X_{lm}$ in
Eq.~(\ref{eq:estimator}) requires inverse weighting the
temperature and polarization data with the full covariance matrix
for the anisotropic survey. This has been done for the WMAP
temperature data at its native resolution~\cite{Smith07,Hanson09}
but extending this to polarization and the resolution required for
Planck requires further work. In practice, fast estimators can
still be constructed for anisotropic surveys, with moderate loss
of performance, by replacing $\bar{a}^X_{lm}$ with some
heuristically-weighted pseudo multipoles (i.e.\ those computed
directly on the masked sky) following techniques used for CMB
power spectrum estimation (e.g.\ see~\cite{Efstathiou04} for a
review). In either case, care must be taken to subtract the
mean-field response -- that obtained on average for no primordial
power asymmetry -- from the quadratic estimator since this is no
longer confined to the $L=0$ mode for an anisotropic survey. The
mean-field and the estimator normalisation are then generally best
determined by Monte-Carlo simulations.

\begin{acknowledgments}
The authors acknowledge use of the Healpix package and thank
Antony Lewis and Duncan Hanson for helpful discussions.
\end{acknowledgments}


\clearpage


\begin{thebibliography}{99}
\bibitem{Bennett03} C. L. Bennett et al., ApJS, \textbf{148}, 1 (2003), astro-ph/0302207.

\bibitem{Hinshaw07} G. Hinshaw et al., ApJS, \textbf{170}, 288 (2007), astro-ph/0603451.

\bibitem{Brown09} M.L. Brown et al., ApJ, \textbf{705}, 978 (2009), arXiv:0906.1003 [astro-ph].

\bibitem{Reichardt09} C.L. Reichardt et al., ApJ, \textbf{694}, 1200 (2009), arXiv:0801.1491 [astro-ph].

\bibitem{ACT10} The ACT Collaboration: J.W. Fowler et al., ApJ 722 (2010) 1148 , arXiv:1001.2934 [astro-ph].

\bibitem{Hall10} N.R. Hall et al., ApJ, \textbf{718}, 632 (2010), arXiv:0912.4315 [astro-ph].


\bibitem{Tegmark03} M. Tegmark, A. de Oliveira-Costa and A. J. S. Hamilton, Phys.
Rev. D, \textbf{68}, 123523 (2003), astro-ph/0302496.

\bibitem{Bielewicz04} P. Bielewicz, K. M. Gorski and A. J. Banday, Mon. Not.
Roy. Astron. Soc. \textbf{355}, 1283 (2004), astro-ph/0405007.

\bibitem{Copi06} C. J. Copi, D. Huterer, D. J. Schwarz and G. D. Starkman,
Mon. Not. Roy. Astron. Soc. \textbf{367}, 79 (2006),
astro-ph/0508047.

\bibitem{Land05} K. Land and J. Magueijo, Phys. Rev. Lett. \textbf{95},
071301 (2005), astro-ph/0502237.

\bibitem{Hansen08} F. K. Hansen et al., ApJ 704 (2009) 1448, arXiv: 0812.3795 [astro-ph].

\bibitem{Eriksen04} H. K. Eriksen et al., Astrophys. J. \textbf{605}, 14
(2004), astro-ph/0307507.

\bibitem{Cruz06} M. Cruz et al., ApJ \textbf{655}, 11 (2007),
astro-ph/0603859.

\bibitem{Cruz07} M. Cruz et al., Science \textbf{318}, 1612 (2007), arXiv:
0710.5737.

\bibitem{Bennett10} C. L. Bennett et al., ApJS \textbf{192} (2010) 17, arXiv: 1001.4758 [astro-ph].

\bibitem{Copi10} C. J. Copi et al., Adv. Astron. Astrophys. \textbf{2010} 2010 847541, arXiv: 1004.5602 [astro-ph].

\bibitem{Ackerman07} L. Ackerman, S. M. Carroll and M. B. Wise, Phys. Rev. D
\textbf{75}, 083502 (2007), astro-ph/0701357.

\bibitem{Himmetoglu09a} B. Himmetoglu, C.R. Contaldi, and M. Peloso, Phys.
Rev. D \textbf{79}, 063517 (2009), arXiv:0812.1231 [astro-ph].

\bibitem{Himmetoglu09b} B. Himmetoglu, C.R. Contaldi, and M. Peloso, Phys.
Rev. Lett. \textbf{102}, 111301 (2009), arXiv: 0809.2779
[astro-ph].

\bibitem{Erickcek08} A. L. Erickcek, M. Kamionkowski and S. M. Carroll,
Phys. Rev. D. \textbf{78}, 123520 (2008), arXiv: 0806.0377
[astro-ph].

\bibitem{Erickcek09} A. L. Erickcek, C.M. Hirata, and M. Kamionkowski Phys.
Rev. D. \textbf{80}, 083507 (2009), arXiv:0907.0705 [astro-ph].

\bibitem{Gordon05} C. Gordon, W. Hu, D. Huterer and T. Crawford, Phys. Rev.
D \textbf{72}, 103002 (2005), astro-ph/0509301.


\bibitem{Dvorkin07} C. Dvorkin, H. V. Peiris and W. Hu, Phys. Rev. D \textbf{77}, 063008 (2008),
arXiv: 0711.2321 [astro-ph].

\bibitem{Groeneboom08} N. E. Groeneboom and H. K. Eriksen, ApJ 690 (2009) 1807, arXiv: 0807.2242
[astro-ph].

\bibitem{Hanson09} D. Hanson and A. Lewis, Phys. Rev. D, \textbf{80}, 063004
(2009), arXiv: 0908.0963 [astro-ph].

\bibitem{Groeneboom09} N. E. Groeneboom, L. Ackerman, I. K. Wehus, and H. K.
Eriksen, ApJ 722 (2010) 452, arXiv: 0911.0150 [astro-ph].

\bibitem{Hanson10} D. Hanson, A. Lewis, and A. Challinor, Phys. Rev. D.,
\textbf{81}, 103003 (2010), arXiv:1003.0198 [astro-ph].

\bibitem{PC05} The Planck Collaboration, 2005, \textit{`The Scientific
Programme of Planck'}, eds. G. Efstathiou, C. Lawrence, and J.
Tauber, ESA-SCI(2005), ESA Publications.

\bibitem{Huffenberger10} K.M. Huffenberger, B.P. Crill, A.E. Lange, K.M.
Gorski, C.R. Lawrence, Astron. Astrophys. 510, A58, 2010, arXiv:
1007.3468 [astro-ph].

\bibitem{Pullen07} A. R. Pullen and M. Kamionkowski, Phys. Rev. D \textbf{76}
103529 (2007), arXiv:0709.1144 [astro-ph].

\bibitem{Gorski05} K. Gorski, E. Hivon, A. Banday, B. Wandelt, F. Hansen, M.
Reinecke, M. Bartelman, ApJ, 622, 759 (2005), astro-ph/0409513.

\bibitem{Hu97} W. Hu and M. White, New Astron. Rev., 2, 323 (1997), astro-ph/9706147.

\bibitem{Zaldarriaga97} M. Zaldarriaga, Phys. Rev. D., \textbf{55}, 1822 (1997), astro-ph/9608050.

\bibitem{Smith07} K. M. Smith, O. Zahn, O. {Dor{\'e}}, Phys. Rev. D., \textbf{76},
043510 (2007), arXiv:0705.3980 [astro-ph].

\bibitem{Efstathiou04} G. Efstathiou, Mon. Not.
Roy. Astron. Soc., 349, 603 (2004), astro-ph/0307515.


\end{thebibliography}
\end{document}